\documentclass{aa}

\usepackage{graphicx}

\usepackage{txfonts}

\begin{document}

\title{Truncated stellar discs in the near infrared. II. Statistical properties and interpretation}

\author{E. Florido\inst{1}, E. Battaner\inst{1}, A. Guijarro\inst{1,2}, F. Garz\'on\inst{3,4} \and A. Castillo-Morales\inst{1}}

\offprints{E. Florido (estrella@ugr.es)}

\institute{Departamento de F\'{\i}sica Te\'orica y del Cosmos, Universidad de Granada, Spain.
   \and
      Centro Astron\'omico Hispano Alem\'an, Almer\'{\i}a, Spain 
   \and
      Instituto de Astrof\'{\i}sica de Canarias, 38200 La Laguna, Spain
   \and
      Departamento de Astrof\'{\i}sica, Universidad de La Laguna, Tenerife, Spain }

\date{}

\abstract{
The results obtained in paper I are used to study possible relationships between the truncation radius of stellar discs in the NIR and structural parameters of the galaxies.
The NIR truncation radius is larger for brighter galaxies, being proportional to  $V_m^c$ with $c \approx 3/2$, and with $V_m$ being the asymptotic rotation velocity at large radii (when the rotation curve becomes flat), and is lower for higher wavelengths. When it is normalized to the scalelength, the truncation is an increasing function of the central surface brightness and is lower for late type galaxies, although these correlations are weaker.
These relations are in agreement with the scenario of magnetically driven truncations.
}
\maketitle   

\section{Introduction}

Paper I (Florido et al. 2005) presented the result of the observation of 18 edge-on spiral galaxies in the NIR in order to study the truncation of the stellar disc, the main objective being to describe the truncation curve. This curve is defined as the progressive decrease of the photometric profile taking place at large radii with respect to the mean exponential profile obtained at moderate radii. In this paper we study the relationship of truncation to other parameters of the galaxy. We investigate some key properties that could restrict some of the theoretical scenarios that have been proposed: what types of galaxy show truncation and what relation exists between the truncation, radius, mass, luminosity, rotation velocity at large radii, type, central surface brightness and  age of stellar population. The results obtained from this study will be merged with those from a previous one (Florido et al. 2001) in order to obtain a larger sample. No merger with other observations in the optical has been attempted, as the assumption of the same behaviour for optical and NIR truncations is not theoretically justified.
    
In early works by van der Kruit (1979) truncations were found to have sharp cut-offs. In paper I and in Florido et al. (2001) truncations were found to be smooth. In these latter two papers it was shown that the smooth behaviour of truncations was not due to projection effects, but was real, existing even in deprojected profiles. In some recent papers (e.g. Pohlen et al. 2002, P\'erez 2004) this description is replaced by a two-exponential profile with a sharp elbow. There is increasing evidence favouring the two-slope description, at least in the optical. We are not interested here in the shape of the profile (see paper I); it may even be that the differences between authors are due to a true difference between the NIR and the optical. For the present, it is not our goal to determine whether this discrepancy is real or is a result of the low signal/noise at these very large radii, especially important in the NIR, and we refer to a ``truncation (break)'' or simply ``truncation''.

In paper I it was proposed that most galaxies in the NIR seem to have a truncation curve that is proportional to $(R-R_{tr})^{-n}$ where $n \approx 1$, $R$ is the radius and $R_{tr}$ is the truncation radius at which the truncation seems to be complete, as obtained by extrapolation of the data.

\section{Theoretical models}

There are at present at least four alternatives to explain truncations (or breaks): a) the collapse model, b) the threshold model, c) the magnetic model and d) the interaction model.

According to the collapse model, which was introduced by van der Kruit (1987), truncations were formed in the collapse of the protogalaxy, and would correspond to the maximum angular momentum of the protogalaxy. The gas beyond the truncation was accreted in later phases of galactic evolution. Ferguson and Clark (2001) presented a model in which the truncation/break is also a result of the initial conditions in the gas.

The threshold model was introduced by Kennicutt (1989). Here, when the gas density falls below a certain threshold, star formation should cease. A recent model by Schaye (2004) improved the dynamic mechanism by which the star formation threshold is established.Van den Bosch (2001) presented a hybrid model combining the collapse and threshold models. Elmegreen and Hunter (2006) reproduce two-slope profiles in a threshold model which includes several local effects of SF.

Kregel and van der Kruit (2004) have shown that the threshold model is better supported by the observations than is the collapse model. The model by Schaye (2004) predicts an increase of $R_{tr}/h$ for small radial scale lengths and an increase of $R_{tr}/h$ for higher central intensities and for low surface brightness discs.

For an extensive discussion of the different models, see the review by Pohlen et al. (2004). See also the discussion in Battaner et al. (2002).

In the magnetic model of truncations (Battaner et al. 2002), a centripetal magnetic force acting on the gas makes an important contribution to the redistribution of stars once they are born.
If the gas rotates under the equilibrium of gravity plus magnetic forces versus the centrifugal force, and if the magnetic force is suddenly suppressed in the star formation process, the newborn stars would migrate and eventually escape. Some important modification of the mass distribution of the stellar density is therefore unavoidably expected at the region where the magnetic force becomes non negligible in the dynamics of the gas.

The interaction model is only rarely defended (Noguchi \& Ishibashi 1986). Schwarzkopf \& Dettmar (2000) found similar properties of truncations for isolated  galaxies and for interacting galaxies. However, close inspection of the large amount of data provided by these authors reveals some differences. Interacting galaxies have a ratio $R_{tr}(R)/ R_{tr}(K)$ that is always larger than one (R is red, K in the NIR), i.e. the truncation radius is larger at shorter wavelengths (with a mean ratio of about 1.6); however, in non-interacting galaxies, this ratio is closer to unity. Thus interaction effects should not be ruled out.

\section{Results}

Paper I presented new observations of edge-on galaxies observed in the NIR. In all, 18 galaxies were observed, but 5 of them were too noisy, 4 untruncated (within observational limits) and 2 uncertain. Therefore, only 7 galaxies were available to study truncations. Five of them, namely NGC 522 (Sbc), MGC-01-05-047 (Sc), NGC 2862 (SBbc), NGC 3279 (Scd) and NGC 5981 (Sc), were very clearly examples of this phenomenon.  There are also two galaxies in which the truncation is not evident but that can be included in our study. These are NGC 781 (Sab) and NGC 3501 (Sc). We also use three truncated galaxies from Florido et al. (2001): NGC 4013 (Sb), NGC 4217 (Sb) and NGC 6504 (Sbc). Therefore the subsample of truncated galaxies adopted in this study is formed by 10 galaxies, 7 from paper I and 3 from Florido et al. (2001). As we have data for both sides of the galaxies, and measure at least in two filters (J and K$_s$), we have 2x2x10 = 40 points for our statistical subsample.   

However, these 40 data points are not independent. Because of the azimuthal symmetry of the galaxies, the results obtained at both sides should be similar (as, in fact, they are) and therefore correlated. Moreover, the closeness of the filter wavelengths, for J and K$_s$ (for a few galaxies, also H), does not permit us to consider the J and K$_s$ points as really independent, except when considering a differential colour behaviour. The fact that we measure in two filters and on both sides permits us to work with smaller errors, but strictly speaking we have only 10 points for our sample, i.e. the number of galaxies selected. Despite the smallness of the sample, we present the more complete data for truncations in the NIR.

{\it A) Truncation versus Hubble type.}

 At first glance, even for such a small number of galaxies, it seems that truncations are found more often in late type galaxies.

 This is a suggestion and is not based on statistics, as the number of galaxies is too low and many more observations would be required. However, we take note of this potential discrimination, as it could be coherent with other findings shown later. This suggestion should be considered with caution, also, because the selection of our samples was not based on physical criteria, but rather considering appropriate angular size, large galactic latitude and date of observations. Therefore the possibility that a bias may be responsible cannot be ruled out.

We looked for a relation of $R_{tr}(J)/h(J)$ versus T, where $R_{tr}$ is the truncation radius, $h$ is the radial scale length and $T$ is the galaxy type. $J$ and $K_s$ are the filters used (Fig.1). The following tendency was observed: the larger the value of parameter $T$, the smaller that of $R_{tr}(J)/h(J)$ is.  This fact is consistent with the above suggestion that truncation is a phenomenon mainly affecting late-type spirals.

\begin{figure}
%\resizebox{\hsize}{!}{\includegraphics{f1tr2.ps}}
\resizebox{\hsize}{!}{\includegraphics{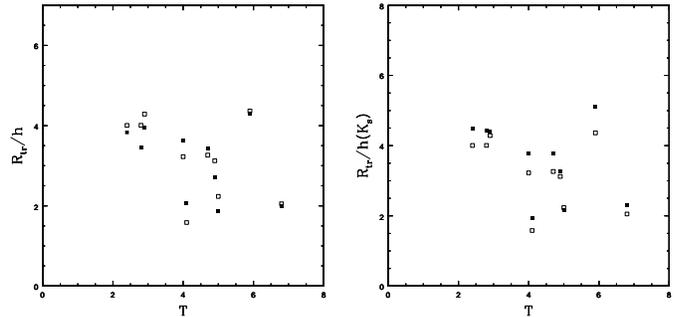}}
\caption{Ratio of truncation radius to scalelength ($R_{tr}/h$) versus morphological type T (from the LEDA database, http://leda.univ-lyon1.fr). Symbols are: open square for $K_s$ and filled square for $J$. In the first panel we plot $R_{tr}/h$ when $R_{tr}$ and $h$ are obtained by the same filter. In the second one we plot $R_{tr}/h(K_s)$ using $h(K_s)$ for this ratio with each filter.}
\label{trun1}
\end{figure}

For some relationships, if we want to adopt $h$ as a length representative of the size of the galaxy, then $h(K_s)$ better represents the ``true'' radial scale length, as it is less affected by extinction and better represents the old stellar population. Therefore, we have plotted $R_{tr}(J)/h(K_s)$ and $R_{tr}(K_s)/h(K_s)$. In this case, the relationship seems to be reinforced.

Therefore, truncation could be a phenomenon that mainly affects late type spirals. This relation is also found when the large sample of Schwarzkopf  \& Dettmar (2000) is used. This suggestion, although attractive, remains to be confirmed.

 \noindent {\it B) Colour dependency of truncations}

Table 1 reveals an evident property: $R_{tr} (J) > R_{tr} (K_s)$. This is so even if the differences are small, because in all galaxies, on all sides, we find $R_{tr} (J) > R_{tr} (K_s)$, with the exception of NGC 6504. Therefore, there is a dependence of $R_{tr}$ with colour. This number of values is not small but we have applied three tests in order to discard an effect of chance: the T test, the Wilcoxon test and the sign test.  In all three tests, the null hypothesis of null difference between $R_{tr} (J)$ and $R_{tr} (K_s)$ was rejected. Taking, as usual, a contrast size of $\alpha =0.05$, in every case the significance level (P-value) was much less than 0.05, being 0.004 for the T test. Therefore,  $R_{tr} (J)$ is unambiguously higher than $R_{tr} (K_s)$. This significant difference informs us about a relation with the age of the population.

\begin{table*}
\centering
\caption[ ]{Basic properties of the galaxies.} 
\begin{flushleft}
\begin{tabular}{|l|r|r|r|r|r|c|c|c|c|} \hline
  Galaxy  &  R$_{tr}$(J)& R$_{tr}$ (K$_s$)& M$_{B}$ & $\log{V_m}$ & T   &  $\mu_o$ (J)& $\mu_o$(K$_s$) &  h (J)  &  h(K$_s$)\\
          &  (Kpc)      &    (Kpc)        &           &             &     &(mag/arcsec$^2$)&(mag/arcsec$^2$)&   (Kpc) &  (Kpc)   \\
    (1)   &  (2)        &    (3)          &    (4)    &    (5)      & (6) &    (7)       &   (8)   &   (9) & (10)   \\
\hline
\hline
NGC 522   &   13.7      &    11.2         &   -20.53  &   2.252     & 4.1 &  20.6        &  21.0  &  6.6 &  7.1  \\
MGC-01-05-047&   29.3   &    27.9         &   -21.77  &   2.410     & 4.9 &  20.9       &  19.6  & 10.8 &  8.9  \\
NGC 781   &   11.0      &     9.8         &   -20.87  &             & 2.4 &  19.6       &  18.1  &  2.9  &  2.4  \\ 
NGC 2862  &   21.7      &    18.6         &   -21.44  &   2.464     & 4.0 &  19.5       &  18.5  &  6.0  &  5.8  \\
NGC 3279  &    7.8      &     6.9         &   -19.27  &   2.208     & 6.8 &  19.8       &  18.4  &  3.9  &  3.4  \\
NGC 3501  &    8.2      &     7.0         &   -19.05  &   2.147     & 5.9 &  20.0       &  18.1  &  1.9  &  1.6  \\
NGC 4013  &    8.5      &     7.8         &   -19.50  &   2.276     & 2.9 &  18.6       &  17.6  &  2.2  &  1.8  \\ 
NGC 4217  &   11.1      &    10.0         &   -19.97  &   2.298     & 2.8 &  18.8       &  17.4  &  3.2  &  2.5  \\
NGC 5981  &   14.6      &    12.6         &   -20.61  &   2.424     & 4.7 &  19.3       &  18.0  &  4.2  &  3.9  \\
NGC 6504  &   22.2      &    23.0         &   -22.53  &             & 5.0 &  19.5       &  18.7  & 11.9  & 10.3  \\
\hline
\end{tabular}

Columns: (1) Galaxy name; (2) Truncation radius in J; (3) Truncation radius in K$_s$; (4) Absolute B-magnitude; (5) Log of maximum velocity rotation (in km/s) from radio observations; (6) Morphological type code; (7) Central surface brightness in J; (8) Central surface brightness in K$_s$; (9) \& (10) Radial scale-lengths in J and K$_s$. Columns 4, 5 and 6 are obtained from the LEDA database (http://leda.univ-lyon1.fr). The values of $R_{tr}(J)$, $R_{tr}(K_s)$, $h(J)$ and $h(K_s)$ were independently obtained for  both sides of each galaxy. The figure given in this table is the mean value. 
\end{flushleft}
\end{table*}

With respect to relative values, taking the radial scale-length as the unit, we obtain $<(R_{tr}/h)>_J = 3.12$ with $\sigma=0.85$ and $<(R_{tr}/h)>_{K_s} = 3.21$ with $\sigma = 0.93$. As mentioned above, $h(K_s)$ better represents the size of the galaxy. Then,  we obtain $<R_{tr}(J)/h(K_s)>= 3.57$ with $\sigma= 1.06$. Of course $<R_{tr}(J)/h(K_s)>$ is significantly higher than $<R_{tr}(K_s)/h(K_s)>$.

These values are in reasonable agreement with other values found in the literature (Pohlen et al. 2002, Barteldrees \& Dettmar 1994). For comparison, it should be taken into account that some authors give $R_{br}$ or the point where the break takes place and others where it ends. Furthermore, our data in the NIR do not have to coincide with optical data, the differences being important.

\noindent {\it C) Truncation and rotation and luminosity.}

Figure 2 clearly shows a relation between $R_{tr}$ and $V_m$, the asymptotic rotation velocity (taken from LEDA). $R_{tr}$ (measured in kpc) is an increasing function of $V_m$, which is emphasized in the log-log plot, showing a relation of the type $R_{tr} \propto V_m^c$. A relationship between $R_{tr}$ and $V_m$ was first obtained observationally by Pohlen et al. (2004). This is clearly supported by our present findings, in another wavelength range, with less scattering and with the mathematical function being determined quantitatively. This fact can be used to restrict the number of theoretical models, as we will see in the discussion.

As different truncation values were detected for J and for K$_s$, we have analyzed this relation ($\log{R_{tr}}$, $\log{V_m}$) for the two filters independently (Fig. 2). Standard Pearson correlations are used because the relation appears to be fairly linear in log space with classical significance tests. For [$\log{R_{tr}}(J)$, $\log{V_m}$] the correlation coefficient is 0.83, the slope is 1.54 with error 0.28 and the $\log{R_{tr}}$ at the origin is -2.45. The P-value is 0.001, much less than 0.05, and therefore the null hypothesis of the absence of a linear relation between both quantities is clearly rejected.
For [$\log{R_{tr}}(K_s)$, $\log{V_m}$] the correlation coefficient is again 0.83, the slope is 1.58 with a typical error of 0.29 and the value of $R_{tr}$ at the origin is -2.61. The P-value is 0.002, again much lower than 0.05.
For both filters, the slope of the regression line (1.54 $\pm$ 0.28 and 1.58 $\pm$ 0.29) can be approximated to the value 1.5, which is more useful when looking for simple low-power relations.

\begin{figure}
\resizebox{\hsize}{!}{\includegraphics{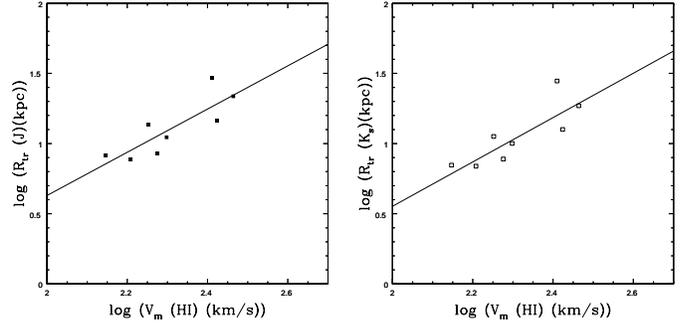}}
\caption{Log of truncation radius versus log of maximum velocity rotation.  Symbols as in Fig. 1}
\label{trun2}
\end{figure}

A relationship physically connected to this one ($R_{tr}$, $V_m$) is the relation ($R_{tr}$, $L$) with $L$ the luminosity of the galaxy. This is shown in Fig. 3, where we plot $\log{R_{tr}}$ versus the absolute B-magnitude, showing a clear relation, with $R_{tr}$ being larger for brighter galaxies. A statistical analysis is unnecessary as this relation is a consequence of the previous one, taking into account the Tully-Fisher relation. 

\begin{figure}
\resizebox{\hsize}{!}{\includegraphics{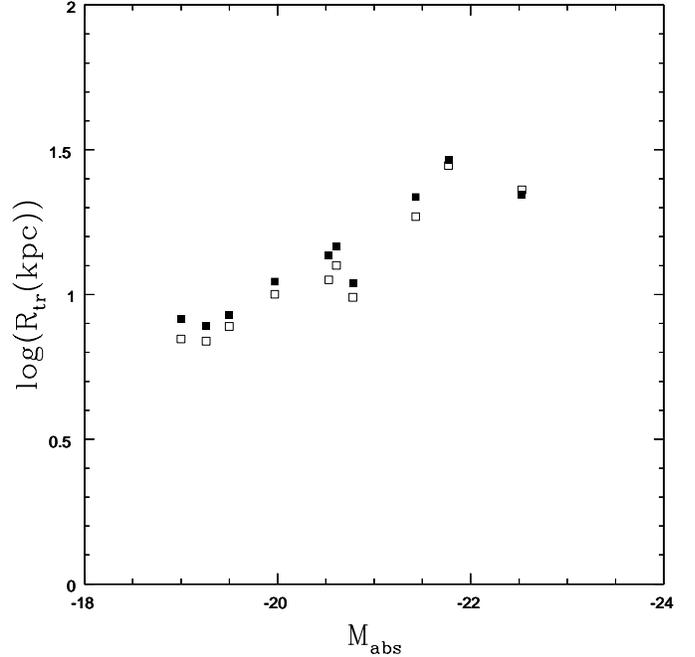}}
\caption{Truncation radius versus absolute B-magnitude. Symbols as in Fig. 1.}
\label{trun3}
\end{figure}

\noindent {\it {\bf D)} Other relationships.}

Fig. 4 plots $R_{tr}$ versus $h$ and shows a clear correlation between the two parameters. Thus, more extended galaxies have larger truncation radii, in agreement with  Kregel (2003).  Larger galaxies have lower truncation radii when measured with the radial scale length as the unit. This relation seems to be non-linear.

Figure 5 shows $R_{tr}/h$ versus $\mu_o$ (deprojected central surface brightness). This relation has been studied by Kregel \& van der Kruit (2004), as mentioned above. Although the scatter of the points is much higher, we see that $R_{tr}/h$ seems to be a decreasing function of $\mu_o$. The correlation coefficient for [$R_{tr}$(J), $\mu_o$] is only 0.39 and is the same for [$R_{tr}$($K_s$), $\mu_o$]. Therefore, the non existence of this relation cannot be statistically rejected, as the P-value is (in both cases) 0.091, larger than the usual value of 0.05. The correlation coefficient for [$R_{tr}/h(K_s)$, $\mu_o$] is again very poor for J (-0.41) and the P-value very large, 0.071 ($>$ 0.05). However, for $K_s$ the P-value is less than 0.04 and the correlation coefficient -0.73. Therefore, for the K$_s$ filter there is a more significant relation: the larger the $\mu_o$ value, the lower the truncation radius when measured with $h(K_s)$ as the unit.

%\begin{figure}
%\resizebox{\hsize}{!}{\includegraphics{f4rtr2.epsi}}
%\caption{Left panel: Truncation radius versus scalelength. Right panel: the same as (a) but always with the filter $K_s$ for the scalelength. Symbols as in Fig. 1.}
%\label{trun4}
%\end{figure}

\begin{figure}
\resizebox{\hsize}{!}{\includegraphics{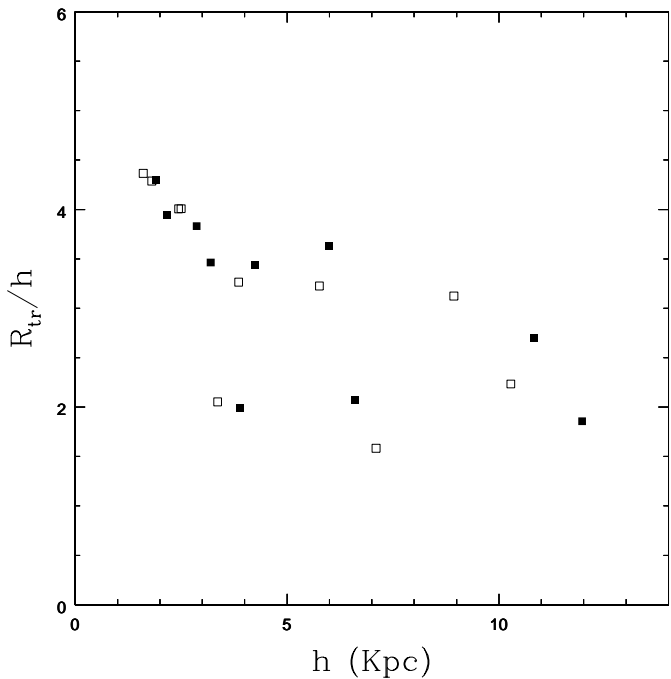}}
\caption{Ratio of truncated radius to scalelength. Symbols as in Fig. 1.}
\label{trun5}
\end{figure}

\begin{figure}
\resizebox{\hsize}{!}{\includegraphics{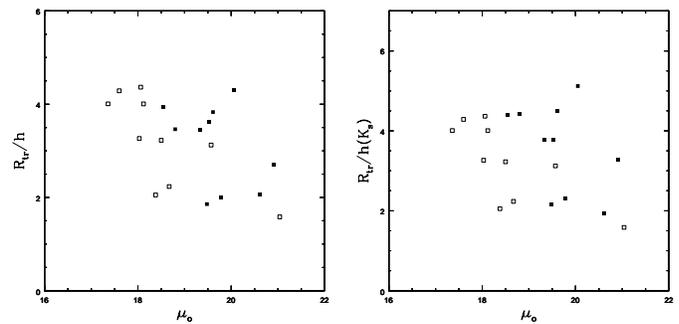}}
\caption{Left panel: Ratio of truncated radius to scalelength versus  deprojected central surface brightness. Right panel: Always considering the scalelength in $K_s$. Symbols as in Fig. 1.}
\label{trun6}
\end{figure}

\section{Discussion}

Despite the relatively small size of our sample (10 galaxies), some clear relationships between the truncation radius and structural parameters of galaxies have been found. These relations are obtained in the near infrared and therefore some caution must be taken when comparing with optical data, as they could differ. For example, star formation can take place beyond the truncation; optical measurements would be sensitive to newborn stars but not NIR measurements. The difference could be very large when comparing data in the U-filter and our NIR data and less so when using the R-band and the I-band. It should also be taken into account that young stars are visible in the NIR, especially red supergiants. Nevertheless, optical and NIR data provide different information. This difference would be larger if stars moved from their birthplace. The results obtained are summarized here:

\begin{itemize}
  \item Late-type galaxies seem to be more affected by the mechanism of truncations.
  
  \item Bright galaxies have larger $R_{tr}$.

  \item $R_{tr}=constant \times V_m^c$ ($c \cong 1.5$). 

  \item Larger galaxies (higher $h$) have larger $R_{tr}$.

  \item $<R_{tr}(J)/h(K_s)> = \alpha(J) \approx 3.57$; $<R_{tr}(Ks)/h(Ks)> = \alpha(Ks) \approx 3.21$; hence, we find lower values for K$_s$. The difference is small but $R_{tr}(J) > R_{tr}(K_s)$ for all galaxies with only one exception. Statistical tests show that this difference is significant.

  \item $R_{tr}/h$ is an increasing function of the central surface brightness, $I_o$, even if the correlation seems to be very weak. 
\end{itemize}

All these properties can be used to put restrictions on the different theoretical models. Other groups should confirm these results with larger samples of galaxies, with independent NIR observations. Clearly, taking into account the fact that some galaxies lie behind many Milky Way stars and/or that the observations were noisy, the number of galaxies in our sample is low and should be enlarged.

The above results agree very well with the magnetic model of Battaner et al. (2002). In this model an inward magnetic force is added to gravity to explain the high rotation velocity and flat rotation curve of the gas.

From the observational point of view, it is now evident that large magnetic centripetal forces exist, as can be directly deduced from observational magnetic field strengths and energy densities in the Milky Way, NGC 6946 and M31 (Beck 2004, Battaner \& Florido 2005). These measurements show large values of the magnetic field strength and, what is more important, such low gradients of the field, i.e. such a low decline of the field for increasing radii, that an important centripetal magnetic force does exist; this is now empirically demonstrated. These magnetic forces act on the gas and are in equilibrium with gravitational and centrifugal forces. When stars are formed this equilibrium is destroyed and stars escape, thus producing the truncation. 

In the simplest version of the magnetic model, Battaner et al. (2002) obtained a formula for the truncation radius

\begin{equation}
   R_{tr} \approx {{2GM} \over {V_m^2}}
\end{equation}
where $M$ is the visible galactic mass and $\theta_0$ is the constant value of the rotation velocity at large radii.

As $M \propto L$ and the Tully-Fisher relation establishes that $L \propto V_m^c$ (where $c$ lies between 3 and 4), it is concluded that
\begin{equation}
  R_{tr} \propto V_m^{(c-2)}
\end{equation}
with the exponent $(c-2)$ being between 1 and 2. This fact matches Fig. 2 very well, showing $R_{tr} \propto V_m^x$, where $x= 1.54 \pm 0.28$ (approximately 1.5 within the estimated error) and the constant of proportionality is 0.004 kpc km$^{-1/2}$ s$^{1/2}$ for J. If we fit the data in $K_s$ the values obtained are: $x= 1.58 \pm 0.29$ and the constant of proportionality is 0.002 kpc km$^{-1/2}$ s$^{1/2}$.  On the other hand, this would imply that more massive (or more luminous) galaxies would have larger $R_{tr}$, so that the truncation might become undetectable if $R_{tr}$ is larger than the last observable point. Using the Tully-Fisher relation again, we find $R_{tr} \propto L^{1-{2 \over c}}$, the exponent ($1-{2 \over c}$) being between 1/3 and 1/2, which also explains Fig. 3.

The magnetic model can also explain the relation between $R_{tr}/h$ and $\mu_0$, the surface brightness at the centre. From equation (2) and $M \propto L \propto I_0 h^2$ (where $L$ is the luminosity and $I_0$ the intensity, in physical units, at the centre) and the Tully-Fisher relation, taking now for simplicity $L \propto V_m^4$
\begin{equation}
  {R_{tr} \over h} \propto {L \over {V_m^2 h}} \propto {L^{1/2}\over h} \propto {{I^{1/2}h} \over h} = I_0^{1/2}
\end{equation}

As $\mu \propto -\log{I}$, the magnetic model predicts a relation between $R_{tr}$ (when measured in radial scale units) and $\mu_0$, the central surface brightness. Any other characteristic galactic magnitude (such as $h$ itself) cancels out.

The magnetic model also predicts $R_{tr}/h \propto M^{1/2}/h$, and hence it explains an observational fact mentioned above: extended galaxies, with large values of $h$, have smaller truncation radii (measured with the radial scale length, $h$, as the unit). In this case, the presence of $M^{1/2}$ in the formula would make the relation $[R_{tr}/h, h]$ be affected by a high degree of scatter, and the relation is certainly more complex than that.

The fact that the magnetic model satisfactorily explains the relations found here does not necessarily imply that all other models should be excluded. For instance, on qualitative grounds, the threshold model (e.g. Schaye, 2004) could predict a relationship between absolute magnitude and truncation if the threshold gas surface density were located further out in a bright galaxy. Some development of this model to explain the findings of this paper is necessary. The magnetic model remains an interesting alternative, as it predicts not only a qualitative relation, but also the mathematical function.

\section{Conclusions} 

The magnetic model satisfactorily explains all the statistical properties reported in this paper. We thus propose this scenario as responsible for producing truncations. The agreement is particularly noticeable concerning the relation between truncation radius and the power 3/2 of the asymptotic rotation velocity.

Magnetic forces are important in the rotation dynamics of the gas, as deduced from observed magnetic field strengths. These forces no longer act when gas is converted into stars, which must have a large influence on the stability of the orbits of young stars, even leading to massive escape.

Large galaxies of type Sa and Sb have less need for dark matter, hence, in the magnetic interpretation, they require lower magnetic fields, i.e. the truncation would be a less important effect in large (massive and luminous) galaxies, especially those with a prominent bulge. The fact that late-type galaxies, which are  bulge-poor, could be more affected by truncations is in agreement with the magnetic scenario. 

Pohlen et al. (2004) found several examples of truncations in edge-on SO galaxies. Even if these galaxies are at present gas-poor, they had both gas and star formation in the past. Thus, the mechanisms suggested by the magnetic hypothesis would have been operating in the past, and they would have inherited truncations.

The fact that the truncation in the optical could be  better described by a two-exponential profile could be compatible with a smooth, complete truncation in the near infrared. The newly formed stars would be, under the magnetic interpretation, in the process of escaping. If the escape speed is roughly equal to $V_m$, i.e. of the order of 100 km s$^{-1}$, and if the length of the escape path is of the order of $\sim$ 10 kpc, the escape time would be of the order of $10^8$ years, a time that differentiates clearly between new-born and old stars. The fact that $R_{tr}(K_s) <R_{tr}(J)$ also fits this interpretation.

In the magnetic interpretation, the truncation corresponds to a true decrease in the total mass (stars + gas) and a connection with a decrease in the rotation velocity should be expected. This decrease has been observed (Casertano 1983, van der Kruit 2001, Bottema 1995) although this fact requires more observational support and may be explained by other means.

\begin{acknowledgements}
This paper has been supported by the ``Plan Andaluz de Investigaci\'on'' (FQM-108) and by the ``Secretar\'{\i}a de Estado de Pol\'{\i}tica Cient\'{\i}fica y Tecnol\'ogica'' (AYA2004-08251-C02-02, ESP2004-06870-C02-02). We are indebted to Andr\'es Gonz\'alez-Carmona who helped us with the statistical analysis. We also acknowledge the LEDA data base organisation, which was highly useful for this research. Thanks to the anonymous referee for valuable comments and suggestions that have contributted to improve the final version of the paper.
\end{acknowledgements}

\end{document}